\documentclass[12pt]{article}
\usepackage{times}
\usepackage{amsmath}
\usepackage{amssymb}
\usepackage{amsthm}
\usepackage{booktabs}
\usepackage{graphicx}
\usepackage{subfigure}
\usepackage[super]{natbib}
\usepackage{natbib}
\usepackage{setspace}
\usepackage{xcolor}
\usepackage[hmargin=1.1in,vmargin=1.1in]{geometry}

\usepackage{atbegshi}
\AtBeginDocument{\AtBeginShipoutNext{\AtBeginShipoutDiscard}}

\newcommand\pr{\normalfont\text{pr}}
\newcommand\w{\wedge}

\begin{document}
\let\cleardoublepage\clearpage

\title{Inference for a test-negative case-control study with added controls}
\author{Bikram Karmakar \\ University of Florida \and Dylan S Small \\ University of Pennsylvania}
\thanks{Address for Correspondence: Bikram Karmakar, Department of Statistics, University of Florida, 226 Griffin Floyd Hall, Gainesville, FL 32611 (E-mail: {\it{bkarmakar@ufl.edu}}).}
\date{\today}
\maketitle

\thispagestyle{empty}


{\it{Running Head}}: Test-negative case-control study with added controls.


{\it{Conflict of interest statement}}: There is no conflict of interest.

{\it{Source of funding}}: None.

{\it{Data and code availability}}: We provide computing code in the supplementary materials.

\newpage

\setcounter{page}{1}

\begin{abstract}
Test-negative designs with added controls have recently been proposed to study COVID-19. An individual is test-positive or test-negative accordingly if they took a test for a disease but tested positive or tested negative. Adding a control group to a comparison of test-positives vs test-negatives is useful since additional comparison of test-positives vs controls can have potential biases different from the first comparison. Bonferroni correction ensures necessary type-I error control for these two comparisons done simultaneously. We propose two new methods for inference which have better interpretability and higher statistical power for these designs.  These methods add a third comparison that is essentially independent of the first comparison, but our proposed second method often pays much less for these three comparisons than what a Bonferroni correction would pay for the two comparisons.
\end{abstract}
{\it keywords}: {Case-control studies; Closed testing; Confidence intervals; Potential biases; Second control group.}

Test-negative studies compare exposures in cases who take a test for a particular disease and test positive vs. controls who also take the test but test negative.\citep{sullivan2016theoretical,vandenbroucke2019test} A test-negative study with added controls (TNSWAC) supplements with controls who did not take the test.  TSNWACs have been used to study antibiotic resistance\citep{sogaard2017risk} and proposed to study COVID-19.\citep{vandenbroucke2020analysis}   The standard inference approach has been to present two exposure rate comparisons,
\begin{itemize}
\item[(i)] test-positives to test-negatives
\item[(ii)] test-positives to controls
\end{itemize}
To control the familywise Type I error rate for multiple comparisons at level $\alpha$ (e.g., $\alpha =0.05$), the Bonferroni inequality can be used and each comparison done at level $\alpha /2$.  Here we propose different inference strategies that can provide greater interpretability and power.

A valuable feature of TSNWACs is that comparisons (i) and (ii) may have different potential biases.\citep{vandenbroucke2020analysis}  Evidence is strengthened when diverse approaches with diverse potential biases produce similar results.\citep{susser1973causal,rosenbaum2010evidence}  However, comparisons (i) and (ii) are dependent {\textendash} see Figure 1A {\textendash} and might tend to agree just because of this dependence.  It is important to distinguish new evidence from the same evidence repeated twice.\citep{rosenbaum2010evidence}
To this end, it is useful to supplement comparisons (i)-(ii) with comparison (iii) test-positives pooled with test-negatives to controls, which is essentially independent of (i) {\textendash} see Figure 1B and supplement {\textendash} and may suffer from different potential biases than (i).\citep{karmakar2020evidence}  For example, it has been hypothesized that smoking protects against Covid-19.\citep{miyara2020}  Comparison (i) may be biased because test-negatives may have some other infection (e.g., the flu) for which smoking increases risk and comparisons (ii) and (iii) may be biased because test-takers tend to be ``health seeking.''\citep{vandenbroucke2019test}  Finding evidence of smoking being protective in all comparisons (i)-(iii) would strengthen evidence compared to just comparisons (i)-(ii) in part because the latter comparisons are dependent.

The following are two procedures that consider comparisons (i)-(iii) and control the familywise error rate for multiple comparisons at $\alpha$ (proof/code in supplement).  The first procedure is (1) test the null hypothesis of no exposure effect in comparison (i), $H_{0(i)}$, at level $\alpha /2$ (i.e., reject if $p$-value $\leq \alpha_2$) and test the null of no exposure effect in comparison (ii), $H_{0(ii)}$, at level $\alpha /2$ and (2) if and only if both nulls are rejected, test the null of no exposure effect in comparison (iii), $H_{0(iii)}$, at level $\alpha$.  The second procedure is
\begin{itemize}
\item[(1)] Test $H_{0(ii)}$ at level $\alpha /2$.  If $H_{0(ii)}$ is rejected, set $\lambda = \alpha$; otherwise, $\lambda = \alpha /2$.
\item[(2)] Test the null of no exposure effect in either comparison (i) and/or comparison (iii), $H_{0(i)}\cup H_{0(iii)}$, at level $\lambda$. This could be done by Fisher's combination method since comparisons (i) and (iii) are essentially independent under the null.\citep{karmakar2020evidence}  If the null $H_{0(i)}\cup H_{0(iii)}$ is not rejected, stop testing.
\item[(3)] If $H_{0(i)}\cup H_{0(iii)}$ was rejected in (2), then test $H_{0(i)}$ and $H_{0(iii)}$ each at level $\lambda$.
\item[(4)] If $\lambda =\alpha /2$ and both $H_{0(i)}$ and $H_{0(iii)}$ were rejected in (3), then test $H_{0(ii)}$ at level $\alpha$ and reject if $p$-value $\leq\alpha$.
\end{itemize}
For example, suppose the p-values for $H_{0(i)}$, $H_{0(ii)}$ and $H_{0(iii)}$ were .04, .03 and .04 respectively, then the standard procedure would not reject any null hypotheses whereas the second procedure would reject all nulls (note: $p$-value for $H_{0(i)} \cup H_{0(iii)}$ using Fisher's combination test is  .012).  Confidence intervals for magnitudes of effect can be formed using both procedures, see supplement.  Figure 1C compares the power of the two proposed procedures and the standard procedure in a simulation.  Both proposed procedures increase power over the standard procedure in the simulated setting with the second procedure providing more power.

\bibliographystyle{plainnat}
\bibliography{test_negative_refs}

\begin{figure}[h!]
\includegraphics[width=\textwidth]{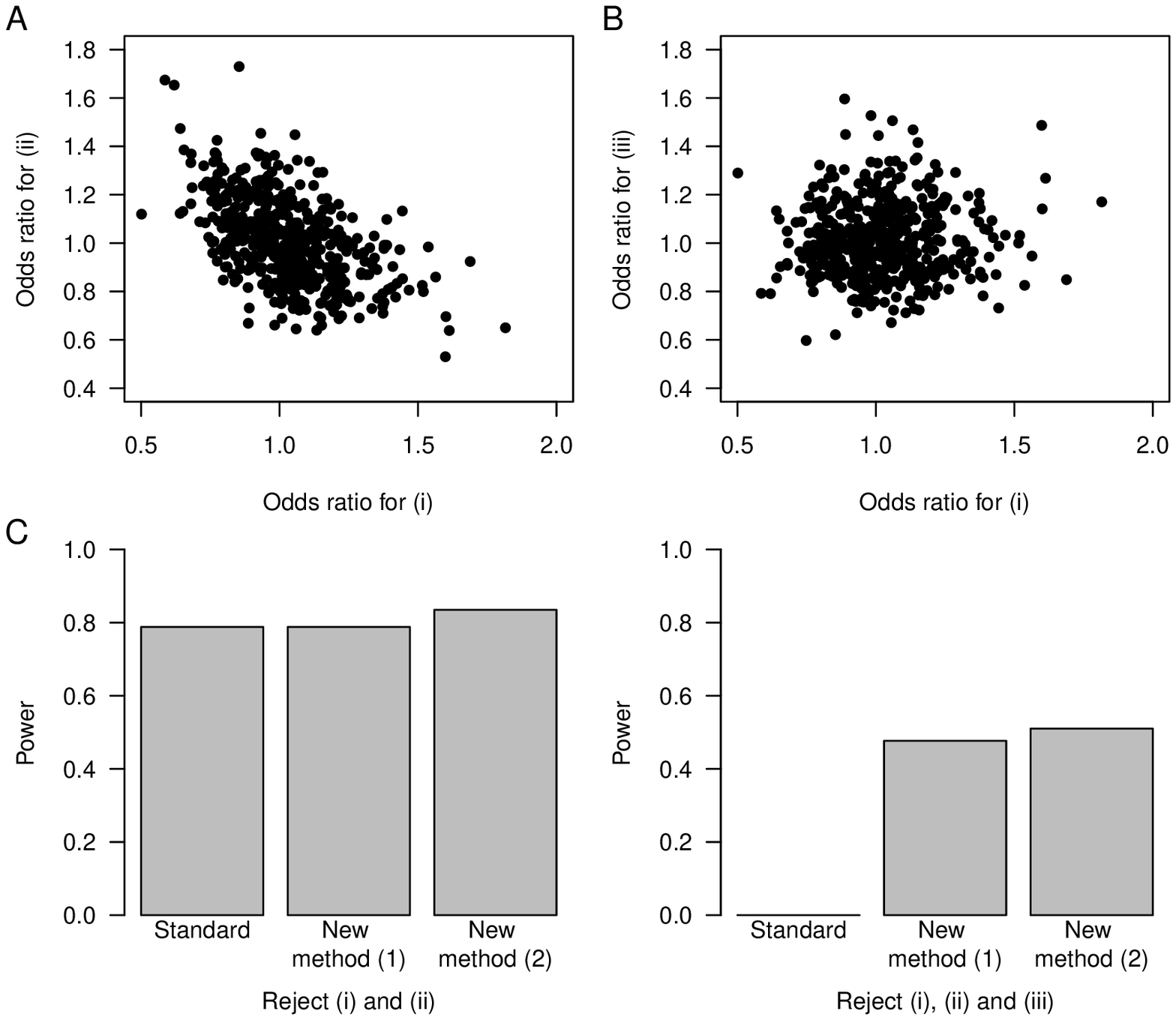}
\label{fig:simulation}
\caption{Simulated results on inference for TNSWAC. In panels A and B the 
null hypothesis is true and there is no difference in the exposure between the test-positives,
test-negatives and controls.  In panel A analyses (i) and (ii) show dependence and in panel B
analyses (i) and (iii) show approximate uncorrelatedness.  Two plots in Panel C show the simulated powers of 
three different methods of analyses of TNSWAC, calculated from 10,000 simulated instances.  
One simulated study consists of 1,250 individuals of, on average, 40\% controls, 30\% 
test-negatives and 30\% test-positives.  Under the null, in Panel A and B, the frequency of an
exposure is constant 20\% in each of these three groups.  In panel C, the odds ratio of an exposure 
is 1.75 for test-positives versus controls and the odds ratio of an exposure is 1.75 for test-positives versus test-negatives. 
R code for the simulation is provided in the supplement.
}
\end{figure}

\newpage

\begin{center}
\newpage
\textsf{\Large  Supplement to ``Inference for a test-negative case-control study with added controls''}
\end{center}
\textsf{Bikram Karmakar and  Dylan Small\let\thefootnote\relax\footnote{Address for Correspondence: Bikram Karmakar, Department of Statistics, University of Florida, 226 Griffin Floyd Hall, Gainesville, FL 32611 (E-mail: {\it{bkarmakar@ufl.edu}}).}\\ 
University of Florida and University of Pennsylvania}

\section{Familywise error rate control}
Setup:  Consider the following three null hypotheses:
$H_{0(i)}$, no difference in exposure between
test-positives and test-negatives; $H_{0(ii)}$, no difference in 
exposure between test-positives and controls; and $H_{0(iii)}$,
no difference in the test-positives or test-negatives and 
controls.  In the following $P_{(i)}$, $P_{(ii)}$ and $P_{(iii)}$
correspond to the three p-values calculated for these 
hypotheses from the corresponding comparisons.

In this setup a method provides a level $\alpha$ 
familywise error rate control if the 
probability of rejecting any true null hypothesis among the three
null hypotheses is at most $\alpha$. In the following we 
let $\mathcal{R}_S$ denote the event that at least one 
of the nulls are rejected among $\{H_{0s} : s\in S\}$
where $S\subseteq \{(i), (ii), (iii)\}$.  We show here that
familywise error rate is controlled for both Method 1 and 
Method 2. \\

\noindent {\bf Method 1.}  Note first that $H_{0(iii)}$ is false 
when and only when one of $H_{0(i)}$ or $H_{0(ii)}$ were false. 

Since Method 1 can reject $H_{0(iii)}$ in step (2) only when both  
$H_{0(i)}$ and $H_{0(ii)}$ are rejected at step (1), we have
$\mathcal{R}_{(iii)} \subseteq \mathcal{R}_{(i)}\cap\mathcal{R}_{(ii)}$, hence
$\mathcal{R}_{(i),(ii),(iii)} \subseteq \mathcal{R}_{(i),(ii)}$. 

To show familywise error rate control, consider now the 
different possibilities of the three hypotheses being true or false separately.

\noindent (a) \ When all three hypotheses are true, the familywise error rate
is 
\begin{align*}
\pr(\mathcal{R}_{(i),(ii),(iii)}) &\leq  \pr(\mathcal{R}_{(i),(ii)})\\
& \leq \pr(\mathcal{R}_{(i)}) + \pr(\mathcal{R}_{(ii)})\\
 &= 
\pr(P_{(i)}\leq \alpha/2) + \pr(P_{(ii)}\leq \alpha/2)\\
 &\leq \alpha/2 + \alpha/2 = \alpha.
\end{align*}

\noindent (b) \ When $H_{0(i)}$ is true but $H_{0(ii)}$ is false, hence $H_{0(iii)}$ 
is false,  the familywise 
error rate is 
$$\pr(\mathcal{R}_{(i)}) = \pr(P_{(i)}\leq \alpha/2) \leq \alpha/2\leq \alpha.$$

\noindent (c) \ Finally, when $H_{0(ii)}$ is true but $H_{0(i)}$ is false, hence $H_{0(iii)}$ 
is false,  the familywise 
error rate is 
$$\pr(\mathcal{R}_{(ii)}) = \pr(P_{(ii)}\leq \alpha/2)\leq \alpha/2\leq \alpha.$$

Hence, the familywise error rate is always controlled.\bigskip

\noindent{\bf Method 2.} \ First we expand the notation $\mathcal{R}_S$ 
to denote the event that at least one of the nulls are rejected among $\{H_{0s} : s\in S\}$
where $S\subseteq \{(i), (ii), (iii), (i)\w (ii)\}$, where
$H_{0(i)\w (iii)} = H_{0(i)}\cap H_{0(iii)}$.  Thus, $H_{0(i)\w (iii)}$ is false is the same as
at least one $H_{0(i)}$ $H_{0(iii)}$ is false, and only when both 
$H_{0(i)}$ and $H_{0(iii)}$ are true we will have $H_{0(i)\w (iii)}$ true. 	

Now we use the result that $P_{(i)}$ and $P_{(iii)}$ are essentially independent and $P_{(i)\w (iii)}$, Fisher's combination of these two p-values, is a valid p-value under $H_{0(i)\w (iii)}$.\cite{karmakar2020evidence} (see footnote)\let\thefootnote\relax\footnote{
Two analyses are essentially independent if the joint distribution of the p-values from these analyses is 
stochastically larger than the uniform distribution on unit square.  Here, (i) and (iii) are nearly independent since we can show 
$\pr(P_{(i)}\leq p, P_{(iii)}\leq q) \leq pq$ for all $0\leq p, q\leq 1$. With larger sample size this inequality becomes
sharper, and asymptotically they are independent.}

Consider again the different combinations of the three hypotheses 
being true or false.  We can reduce some effort in this  enumeration by noting that 
$H_{0(iii)}$ is false when and only when one of $H_{0(i)}$ or $H_{0(ii)}$ 
were false.

\noindent (a) \ When all three of $H_{0(i)}$, $H_{0(ii)}$ and $H_{0(iii)}$ are true, the 
familywise error rate is 
\begin{align*}
\pr(\mathcal{R}_{(i),(ii),(iii)})  & \leq \pr(\mathcal{R}_{(ii)} \text{ at level } \alpha/2 \text{ in step (1) or }
\mathcal{R}_{(i)\w (iii)} \text{ at level } \alpha/2 \text{ in step (2)})\\
&\leq \pr(\mathcal{R}_{(ii)} \text{ at level } \alpha/2) + \pr(\mathcal{R}_{(i)\w (iii)} \text{ at level } \alpha/2)\\
& = \pr(P_{(ii)}\leq \alpha/2) + \pr(P_{(i)\w (iii)} \leq \alpha/2)\\
& \leq \alpha/2 + \alpha/2 = \alpha.
\end{align*}

\noindent (b) \ When $H_{0(i)}$ is true but $H_{0(ii)}$ is false, hence $H_{0(iii)}$ 
is false,   the 
familywise error rate is 
\begin{align*}
\pr(\mathcal{R}_{(i)}) & = \pr(\mathcal{R}_{(i)} \text{ at level } \alpha/2 \text{ or at level } \alpha \text{ in step (3), by whether $\lambda = \alpha/2$ or $=\alpha$})\\
& \leq \pr(\mathcal{R}_{(i)} \text{ at level } \alpha)\\
& =\pr(P_{(i)}\leq \alpha) \leq \alpha.
\end{align*}

\noindent (c) \ Finally, when $H_{0(ii)}$ is true but $H_{0(i)}$ is false, hence $H_{0(iii)}$ 
is false, the familywise error rate is \vspace{-10pt}
\begin{align*}
\pr(\mathcal{R}_{(ii)}) & = \pr(\mathcal{R}_{(ii)} \text{ at level } \alpha/2 \text{ or at level } \alpha \text{ in step (3), by whether $\lambda = \alpha/2$ or $=\alpha$})\\
& \leq \pr(\mathcal{R}_{(ii)} \text{ at level } \alpha)\\
& =\pr(P_{(ii)}\leq \alpha) \leq \alpha.
\end{align*}

Hence, the familywise error rate is always controlled.

\section{Confidence sets for the magnitude of effects}

{\bf Notation:} \ We can create confidence sets for the effects of the exposure using the 
methods discussed in the letter.  Some new notation are needed. 
In the following a subscript $P$ is for test-positives, $N$ for test-negatives,
and $C$ for the added controls.  Also, $n$ with appropriate 
subscript denotes the counts of a particular  group of individuals.
For example, $n_{P1}$ denotes the number of exposed test-positives
and $n_{C0}$ the number of unexposed test-negatives, and 
$n_{PN1}$ is the number of exposed test-positives or test-negatives.\smallskip

\noindent{\bf Data tables:} \ The collected data can be tabulated in three
tables corresponding to the three comparisons (i), (ii) and (iii).\medskip\medskip

\noindent\makebox[\textwidth][c]{
\begin{minipage}{3in}

\begin{tabular}{lcc}
\multicolumn{3}{c}{Comparison (i)}\\
& Exposed & Unexposed\\
Test-positive & $n_{P1}$ & $n_{P0}$\\

Test-negative & $n_{N1}$ & $n_{N0}$\\\hline
Total & $n_{PN1}$ & $n_{PN0}$
\end{tabular}

\end{minipage}\hspace{20pt}
\begin{minipage}{3in}

\begin{tabular}{lcc}
\multicolumn{3}{c}{Comparison (ii)}\\
& Exposed & Unexposed\\
Test-positive & $n_{P1}$ & $n_{P0}$\\

Control & $n_{C1}$ & $n_{C0}$\\\hline
Total & $n_{PC1}$ & $n_{PC0}$
\end{tabular}
\end{minipage}}

\noindent\makebox[\textwidth][c]{

\begin{tabular}{lcc}
\multicolumn{3}{c}{Comparison (iii)}\\
& Exposed & Unexposed\\
Test-positive or negative & $n_{PN1}$ & $n_{PN0}$\\

Control & $n_{C1}$ & $n_{C0}$\\\hline
Total & $n_{PNC1}$ & $n_{PNC0}$\\
\end{tabular}}
\medskip\medskip

A p-value for a given one of the three comparisons can be calculated 
from the corresponding table, e.g., using Fisher's exact test.  
For example, $P_{(ii)}$ is the p-value calculated from the 
2-by-2 table above with the numbers $n_{P1}, n_{C1}, n_{P0}$ and 
$n_{C0}$. \smallskip

\noindent{\bf Effects of interest:} \ We have three effects of interest for the exposure, between
test-positives and test-negatives, between test-positives and controls, and 
one between test-negatives and controls.
We denote these effects as $\theta_{P,N}, \theta_{P,C}$ and $\theta_{N,C}$, which 
are defined below. These are called attributable effects.

The effect $\theta_{P,N}$ is the ratio of the number of individuals who
became test-positive because of the exposure, but in the absence of it 
would have been test-negative minus the number of individuals who became 
test-negative because of the exposure but in the absence of it would have been 
test-positive, divided by the number of exposed test-positives or test-negatives.
Notice that $\theta_{P,N}$ is a number between -1 and 1;
$\theta_{P,N}=0$ if exposure did not move anyone from being test-positive compared
to test-negative without exposure or the reverse. 
If $\theta_{P,N}$ is positive, there individuals for whom the exposure
caused them to become test-positive.  Similarly, if $\theta_{P,N}$ is negative,
there are individuals for whom the exposure caused them to become test-negative.
In summary, $\theta_{P,N}$is the net effect of the exposure on becoming
test-positive over test-negative for exposed tested individuals.
%

The second effect $\theta_{P,C}$ is defined similarly. By our definition, $\theta_{P,C}$ 
is the net effect of the exposure for test-positives versus controls relative to
 all exposed individuals either test-positive or control. We have $\theta_{P,C} = 0$ 
 if the exposure did not make any change in who became test-positive over control or 
 the reverse.
 
Finally, we define a third attributable effect $\theta_{N,C}$ in the same way to denote
the net effect of the exposure on becoming test-negative over control for  all exposed
 non test-positive individuals.


A method that calculates p-values using the three tables above is testing the
hypothesis of no effect of the exposure that $\theta_{P,N} = 0, \theta_{P,C}=0$ and $\theta_{N,C}=0$.\smallskip
 
\noindent{\bf  Confidence sets:} \ We construct confidence sets for the effects $\theta_{P,N}, \theta_{P,C}$ and $\theta_{N,C}$.  To do this we have to explain how to test that 
 $\theta_{P,N}=\theta_{P,N}^{^\star}, \theta_{P,C}=\theta_{P,C}^{^\star}$ and $\theta_{N,C}=\theta_{N,C}^{^\star}$ where 
 $\theta_{P,N}^{^\star}, \theta_{P,C}^{^\star}$ and $\theta_{N,C}^{^\star}$ could be different from 0, not no effect of the  exposure.  When they are
 different from 0, we adjust the observed tables based on these effects to create tables of the 
 potential outcomes under no exposure.\medskip
 
\noindent\makebox[\textwidth][c]{
\begin{tabular}{lcc}
\multicolumn{3}{c}{Adjusted comparison (i)}\\
& Exposed & Unexposed\\
Test-positive & $n_{P1} - \theta_{P, N}^{^\star}n_{PN1} - \theta_{P, C}^{^\star}n_{PC1}$ & $n_{P0}$\\

Test-negative & $n_{N1} + \theta_{P, N}^{^\star}n_{PN1} - \theta_{N,C}^{^\star}n_{NC1}$ & $n_{N0}$\\
\end{tabular}}\medskip

\noindent\makebox[\textwidth][c]{
\begin{tabular}{lcc}
\multicolumn{3}{c}{Adjusted comparison (ii)}\\
& Exposed & Unexposed\\
Test-positive & $n_{P1} - \theta_{P, N}^{^\star}n_{PN1} - \theta_{P, C}^{^\star}n_{PC1}$ & $n_{P0}$\\

Control & $n_{C1} + \theta_{P, C}^{^\star}n_{PC1}+\theta_{N,C}^{^\star}n_{NC1}$ & $n_{C0}$\\
\end{tabular}}\medskip

\noindent\makebox[\textwidth][c]{
\begin{tabular}{lcc}
\multicolumn{3}{c}{Adjusted comparison (iii)}\\
& Exposed & Unexposed\\
Test-positive or negative& $n_{PN1} - \theta_{P, C}^{^\star} n_{PC1}-\theta_{N,C}^{^\star}n_{NC1}$ & $n_{PN0}$\\

Control & $n_{C1} + \theta_{P, C}^{^\star}n_{PC1}+\theta_{N,C}^{^\star}n_{NC1}  $ & $n_{C0}$\\
\end{tabular}}
\medskip\medskip

Using either Method 1 or Method 2 we could test these three tables at level $\alpha$. 
Either method will make decisions to reject or not reject these adjusted tables. 
Then we write ${R}_{(i)}(\theta_{P,N}^{^\star}, \theta_{P,C}^{^\star},  \theta_{N,C}^{^\star}),$
${R}_{(ii)}(\theta_{P,N}^{^\star}, \theta_{P,C}^{^\star},  \theta_{N,C}^{^\star})$ and
 ${R}_{(iii)}(\theta_{P,N}^{^\star}, \theta_{P,C}^{^\star},  \theta_{N,C}^{^\star})$ as 
 binary variables which are 1 or 0 according to whether comparison (i), 
 (ii) or (iii) is rejected, respectively, based on these adjusted
 tables.  Our confidence interval is
 $$\Big\{ (\theta_{P,N}^{^\star}, \theta_{P,C}^{^\star},  \theta_{N,C}^{^\star}) : \prod_{s\in \{(i), (ii), (iii)\}}{R}_{s}(\theta_{P,N}^{^\star}, \theta_{P,C}^{^\star},  \theta_{N,C}^{^\star})=0\Big\}.$$
 
 Since either method performed at level $\alpha$ provides familywise error rate control at $\alpha$, 
 this confidence interval will have a minimal coverage of $1-\alpha$ for both Method 1 and Method 2.

\section{\protect{\texttt{R}} code to implement new method (2)}

Let \verb|p_i|, \verb|p_ii| and \verb|p_iii| be variables in \texttt{R}
that record the p-values from the three comparisons.  They can be 
calculated using the syntax 
\verb|p_i = fisher.test(e_i, g_i)$p| where \verb|e_i| is a variable 
recording of exposure status, and \verb|g_i| is a variable recording the
case status only for the test-positives and test-negatives.   \verb|e_ii|, 
 \verb|g_ii| and  \verb|e_iii|, \verb|g_iii| have the same role in the
 following code corresponding to the comparisons (i) and (iii) respectively.

\begin{verbatim}
## Significance level for familywise error rate control
alpha <- 0.05
alpha.2 <- alpha/2

## p-values computed from the three comparisons
p_i = fisher.test(e_i, g_i)$p
p_ii = fisher.test(e_ii, g_ii)$p
p_iii = fisher.test(e_iii, g_iii)$p

### Start of Method 2	###
r_i = r_ii = r_iii = 0 		# an inference for reject, value 1, or 0.
## Step (1)
r_ii = 1*(p_ii < alpha.2)
lambda = ifelse(r_ii, alpha, alpha-alpha.2)
## Step (2)
# Fisher's combination 
p_i_or_iii = pchisq(-2*log(p_i*p_iii), 4, lower.tail=FALSE)
r_i_or_iii = 1*(p_i_or_iii < lambda)
## Step (3)
if(r_i_or_iii)
     r_i = 1*(r_i < lambda);    r_iii = 1*(r_iii < lambda)
## Step (4)	
if(r_i & r_iii)  r_ii = 1*(p_ii < alpha)
### Final inference
c(r_i, r_ii, r_iii)
#### END OF CODE ####
\end{verbatim}

%
%
%
%
%

\end{document}